\documentclass{article}

     \usepackage[final]{neurips_2019}

\usepackage[utf8]{inputenc} 
\usepackage[T1]{fontenc}    
\usepackage{hyperref}       
\usepackage{url}            
\usepackage{booktabs}       
\usepackage{amsfonts}       
\usepackage{nicefrac}       
\usepackage{microtype}      
\usepackage{graphicx}
\usepackage{amsmath}
\usepackage{natbib}

\title{Adaptive-CS-Net: \\ FastMRI with Adaptive Intelligence}

\author{%
  Nicola~Pezzotti\thanks{denotes equal contribution.}\\
  Philips Research\\
  Eindhoven, The Netherlands \\
  \texttt{nicola.pezzotti@philips.com} \\
  \And
  Elwin de~Weerdt\footnotemark[1] \\
  Philips Healthcare \\
  Best, The Netherlands \\
  \texttt{elwin.de.weerdt@philips.com} \\
      \And
  Sahar Yousefi\footnotemark[1] \\
  Leiden University Medical Center \\
  Leiden, The Netherlands \\
  \texttt{s.yousefi.radi@lumc.nl}\\
    \And
  Mohamed S. Elmahdy\footnotemark[1] \\
  Leiden University Medical Center \\
  Leiden, The Netherlands \\
  \texttt{m.s.e.elmahdy@lumc.nl}\\
  \AND
  Jeroen~van~Gemert\footnotemark[1]\\
  Philips Healthcare\\
  Best, The Netherlands\\
  \texttt{jeroen.van.gemert@philips.com}\\
  \And
  Christophe Sch\"ulke \\
  Philips Research \\
  Hamburg, Germany \\
  \texttt{christophe.schuelke@philips.com}\\
    \And
  Mariya Doneva \\
  Philips Research \\
  Hamburg, Germany \\
  \texttt{mariya.doneva@philips.com}\\
    \And
  Tim Nielsen \\
  Philips Research \\
  Hamburg, Germany\\
  \texttt{tim.nielsen@philips.com}\\
    \And
  Sergey Kastryulin \\
  Philips Research \\
  Moscow, Russia \\
  \texttt{sergey.kastryulin@philips.com}\\
      \And
  Boudewijn P.F. Lelieveldt\\
  Leiden University Medical Center \\
  Leiden, The Netherlands \\
  \texttt{b.p.f.lelieveldt@lumc.nl}\\
    \And
  Matthias J.P. van Osch\\
  Leiden University Medical Center \\
  Leiden, The Netherlands \\
  \texttt{m.j.p.van\_osch@lumc.nl}\\
    \And
  Marius Staring \\
  Leiden University Medical Center \\
  Leiden, The Netherlands \\
  \texttt{m.staring@lumc.nl}\\
}

\begin{document}

\maketitle

\begin{abstract}
Adaptive intelligence aims at empowering machine learning techniques with the extensive use of domain knowledge.
In this work, we present the application of adaptive intelligence to accelerate MR acquisition. 
Starting from undersampled k-space data, an iterative learning-based reconstruction scheme inspired by compressed sensing theory is used to reconstruct the images. 
We adopt deep neural networks to refine and correct prior reconstruction assumptions given the training data.
Our results show that an adaptive intelligence approach performs better than traditional methods as well as deep learning methods that do not take prior knowledge into account.
\end{abstract}

\section{MR reconstruction background: compressed sensing}
The current state-of-the-art of the commercial MRI reconstruction products is based on forms of compressed sensing (CS) theory, where (a form of) the following optimization problem is solved:
\begin{equation}
\min_{\mathbf{x}} \lbrace F(\mathbf{x}) \equiv \parallel \mathbf{A} \mathbf{x} - \mathbf{b} \parallel ^{2} _{2} + \lambda \parallel \mathbf{R}\mathbf{x} \parallel_{1} \rbrace,
\label{eq:basicL2L1}
\end{equation}
where the $\mathrm L_1$ norm is used to enforce sparsity of the solution in a domain specified by the transformation $\mathbf{R}$ such as wavelet transform. The optimization problem is convex and can be solved using a method in the class of Iterative Shrinkage-Thresholding Algorithms (ISTA). Recently, Zahng et. al. proposed a non-linear transformation using a deep learning network called ISTA-Net \cite{zhang2018ista}. Our work leverages this approach by including domain specific prior knowledge.

\section{Adaptive-CS-Net}
\begin{figure}
\begin{center}
\includegraphics[width=0.8\textwidth]{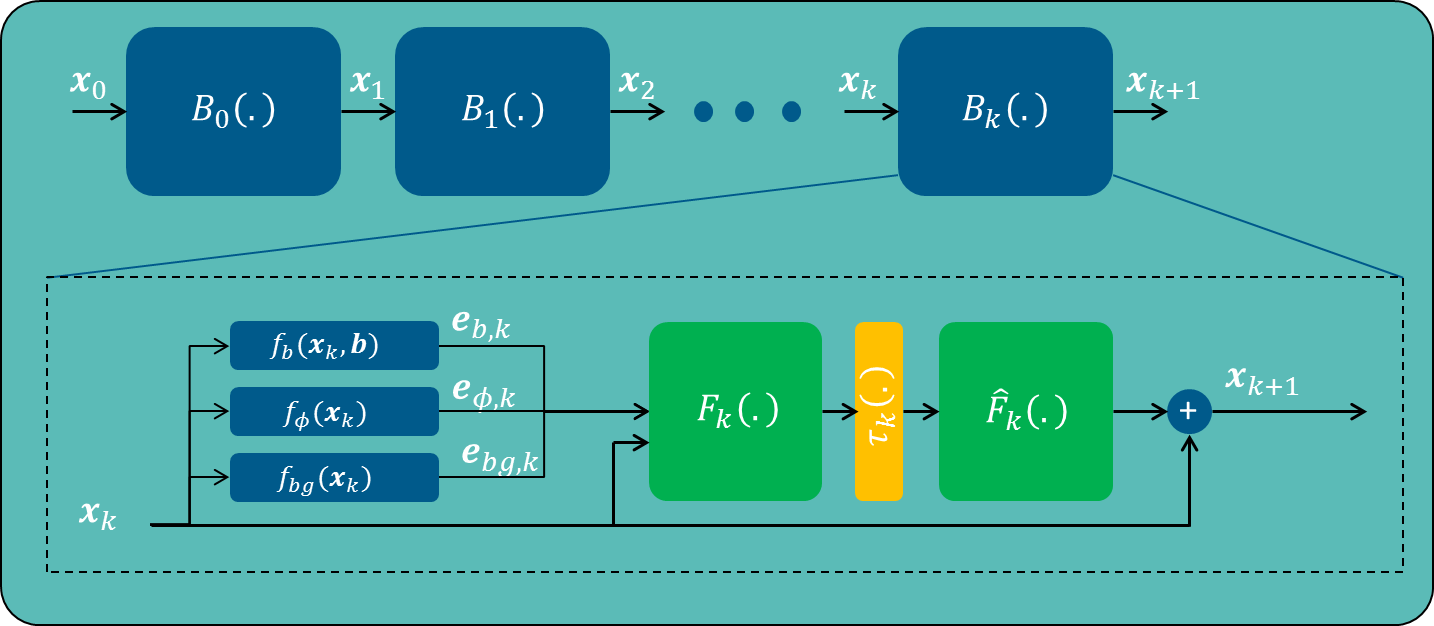}
\caption{Proposed adaptive Adaptive-CS-Net architecture}
\label{fig:our_solution_1}
\end{center}
\end{figure}

Figure \ref{fig:our_solution_1} illustrates the proposed solution, in which the general update step is defined by:
\begin{equation}
\mathbf{x}_{k+1} = B_k(\mathbf{x}_k) = \mathbf{x}_k + \hat{F_k} \left(\tau_{\lambda_k} \left( F_k \left( \mathbf{x}_k, \mathbf{e}_{b,k}, \mathbf{e}_{\phi,k}, \mathbf{e}_{bg,k} \right) \right)  \right), 
\label{eq:our_net_1}
\end{equation}
where $F_k(.)$ and $\hat{F}_k(.)$ are neural networks. 
This approach is similar to the one presented by Zhang \& Ghanem \cite{zhang2018ista}, but we employ UNet-shaped networks \cite{ronneberger2015u} as sparsifying transformer to exploit multi-scale information during the reconstruction. 
The function $F_k(.)$ represents the encoder path of the UNet, $\tau_{\lambda_k}$ the learnable soft thresholding function applied to the features maps before the skip connection and, finally, $\hat{F}_k(.)$ represents the decoder path of the UNet.
Besides the image $\mathbf{x}_k$, input to the network are domain specific information (priors) $\mathbf{e}_{b,k}$, $\mathbf{e}_{\phi,k}$ and $\mathbf{e}_{bg,k}$. These domain knowledge terms represent soft data consistency $\mathbf{e}_{b,k}$, known phase behavior given the spin-echo behavior $\mathbf{e}_{\phi,k}$, and the location of the background as obtained from a low frequency reconstruction as a regularization $\mathbf{e}_{bg,k}$.

In order to account for the strong correlation between neighbouring slices, we adopt a 2.5D learning approach. 
In this approach, we reconstruct the target slice and its neighboring slices simultaneously. The vector $\mathbf{x}_k$ can then be seen as the concatenation of the set of slices.

A weighted sum of Multiscale-SSIM (MSSIM) \cite{wang2003multiscale} and $\mathrm L_1$ was used as the loss function:
\begin{equation}
Loss = \alpha \text{MSSIM}( \mathbf{t} , \mathbf{x}^c ) + (1-\alpha) \Vert \mathbf{t} - \mathbf{x}^c \Vert_1,
\end{equation}
where $\mathbf{t}$ is the target image, $\mathbf{x}^c$ is the reconstructed center image, and $\alpha$ is 0.84. Note that the loss is applied only on the center slice and the network is free to learn to reconstruct the neighboring slices only if it helps to better reconstruct $\mathbf{x}^c$. 
MSSIM encourages the network to learn and preserve structural and contrast information, while $\mathrm L_1$ enforces sensitivity to uniform biases to preserve luminance \cite{zhao2015loss}.    

\subsection{Network training and implementation details}
Two models consisting of 25 blocks in the ISTA-Network were trained for the single- and multi-coil datasets provided by the fastMRI challenge \cite{zbontar2018fastMRI} for 25 epochs. These models were trained on accelerations ranging from 2x to 10x. These models are subsequently fine-tuned for 10 epochs for only 4x and 8x accelerations and protocol (PD and PDFS).    
The networks, which contain 33M trainable parameters each, were trained on two RTX8000 NVIDIA cards using Pytorch \cite{paszke2017automatic}. As optimizer, we used the state-of-the-art Rectified Adam (RAdam) introduced by Liu et al. \cite{liu2019variance} using a decaying learning rate of $10^{-4}$ and a batch size of 6.

\section{Conclusion and discussion}
The resulting neural network was trained and tested on data from the FastMRI challenge, showing good quality of the reconstructed images for the single and multi-coil reconstruction using both accelerations. 
Adaptive-CS-Net is declared the best performing solution in the multi-coil 8x accelerated track in the first edition of the FastMRI challenge. 

In this work, we propose Adaptive-CS-Net, an adaptive intelligence approach to MR image reconstruction that incorporates prior information such as data consistency and known phase behaviour. 
Explicitly feeding prior information to each network block constrains the solution, enabling a self-correction mechanism of the iterative estimate made by the network.

\bibliographystyle{unsrt} 
\bibliography{references}

\end{document}